\documentclass[]{aa}
\usepackage{natbib}
\bibpunct{(}{)}{;}{a}{}{,} 
\usepackage{graphicx}
\usepackage{txfonts}

\def\thco{$^{13}$CO}
\def\eico{C$^{18}$O}
\def\meth{CH$_3$OH}
\def\hhco{H$_2$CO}
\def\hh{H$_2$}
\def\cch{C$_2$H}

\def\nmb{$N_{\rm mb}$}

\def\kkms{K\,km\,s$^{-1}$}
\def\kms{km\,s$^{-1}$}
\def\scm{cm$^{-2}$}
\def\ccm{cm$^{-3}$}

\def\pow#1#2{#1$\times$10$^{#2}$}

\def\HII{\hbox{\rm H\,{\sc ii}}\ }

\def\CIInospace{\hbox{\rm C\,{\sc ii}}}
\def\OInospace{\hbox{\rm O\,{\sc i}}}

\begin{document}

\title{Chemical stratification in the Orion Bar: JCMT Spectral Legacy Survey observations}

\author{M.~H.~D.~van der Wiel\inst{1,2}, F.~F.~S.~van der Tak\inst{2,1},
  V.~Ossenkopf\inst{3,2}, M.~Spaans\inst{1}, H.~Roberts\inst{4},
  G.~A.~Fuller\inst{5}, R.~Plume\inst{6}} 
\authorrunning{M.~H.~D.~van der Wiel et al.}
\titlerunning{Chemical stratification in the Orion Bar: JCMT-SLS data}
\institute{
Kapteyn Astronomical Institute, P.O. Box 800, 9700 AV, Groningen, The Netherlands; email: \texttt{wiel@astro.rug.nl}
\and 
SRON Netherlands Institute for Space Research, P.O. Box 800, 9700 AV, Groningen, The Netherlands
\and
I. Physikalisches Institut der Universit\"at zu K\"oln, Z\"ulpicher Stra\ss{}e 77, 50937 K\"oln, Germany
\and
Astrophysics Research Centre, School of Mathematics and Physics, Queen's University of Belfast, Belfast, BT7 1NN, UK
\and
Jodrell Bank Centre for Astrophysics, Alan Turing Building, University of Manchester, Manchester, M13 9PL, UK
\and
Department of Physics and Astronomy, University of Calgary, Calgary, T2N 1N4, AB, Canada
}

\date{Received 21 November 2008 / Accepted 6 February 2009 / Corrected 18 December 2009}

\abstract
{Photon-dominated regions (PDRs) are expected to show a layered structure in molecular abundances and emerging line emission, which is sensitive to the physical structure of the region as well as the UV radiation illuminating it.}
{We aim to study this layering in the Orion Bar, a prototypical nearby
PDR with a favorable edge-on geometry.}
{We present new maps of 2$\arcmin$$\times$2$\arcmin$ fields at 14\arcsec--23\arcsec\ resolution toward the Orion Bar
 in the SO 8$_8$--9$_9$, \hhco\ 5$_{15}$--4$_{14}$, \thco\ 3--2, \cch\
 4$_{9/2}$--3$_{7/2}$ and 4$_{7/2}$--3$_{5/2}$, \eico\ 2--1 and HCN 3--2
 transitions. }
{The data reveal a clear chemical stratification pattern. The \cch\
emission peaks close to the ionization front, followed by \hhco\ and SO, while
\eico, HCN and \thco\ peak deeper into the cloud. A simple PDR model reproduces
the observed stratification, although the SO emission is predicted to peak
much deeper into the cloud than observed while \hhco\ is predicted to peak closer to the ionization front than observed. In addition, the predicted SO
abundance is higher than observed while the \hhco\ abundance is lower than
observed.}
{The discrepancies between the models and observations indicate
that more sophisticated models, including production of \hhco\ through grain surface chemistry,
are needed to quantitatively match the observations of this region.}


\maketitle

\section{Introduction}
\label{sec:intro}

In photon-dominated regions (PDRs), UV radiation between a few and 13.6\,eV drives the
thermal and chemical balance of interstellar gas \citep[and references therein]{hollenbach_tielens1999}.
PDRs are ubiquitous in the universe: surfaces of molecular
clouds adjacent to \HII regions, planetary nebulae, protoplanetary disks and also the nuclei
of distant active galaxies. As such, PDRs are signposts of radiative feedback processes driven
by star formation. Heating proceeds through photo-electric emission from dust
grains and cooling is mostly due to fine-structure emissions of [\CIInospace] and [\OInospace], and CO rotational
lines \citep[e.g.,][]{sternberg1995,kaufman1999,meijerink2005,rollig2007}. 
Shielding of the UV radiation by dust and gas creates a layering structure
where a sequence of different chemical transitions is produced by the gradual attenuation
of the UV field \citep{ossenkopf2007}.

The Orion Bar is a prototypical PDR located between the Orion molecular cloud
and the \HII region surrounding the Trapezium stars, at a distance of 414\,pc
\citep{menten2007}. Multi-wavelength observations
\citep{tielens1993,vanderwerf1996,hogerheijde1995,youngowl2000, walmsley2000}
indicate a geometry for the Bar where the PDR is wrapped around the \HII region
created by the Trapezium stars and changes from a face-on to an edge-on geometry
where the molecular emissions peak. The mean density of the Bar is about $10^5$
cm$^{-3}$, the mean molecular gas temperature 85 K, and the impinging radiation
field is $(1-4)\times 10^4\,\chi_0$ \citep{hogerheijde1995}, where the Draine
field $\chi_0 = 2.7\times 10^{-3}$ erg s$^{-1}$ cm$^{-2}$.

The clumpiness of the PDR inferred by \citet{hogerheijde1995} was confirmed by
interferometric data \citep{youngowl2000, lis2003}. Clump densities
up to $10^7$ cm$^{-3}$ were derived by \citet{lis2003} while the density of
the interclump medium should fall between a few $10^4$ cm$^{-3}$
\citep{youngowl2000} and $2\times 10^5$ cm$^{-3}$ \citep{simon1997}.
The physical stratification of the PDR is well established 
\citep{tielens1993,vanderwerf1996,simon1997,lis2003}. Vibrationally excited H$_2$ emission is located 15$\arcsec$ from the ionization
front, where HCO$^+$ 1--0 peaks as well, while the CO peak resides at 20$\arcsec$
and CS slightly further in.

This paper presents emission maps of various molecular species, allowing us to
probe and understand the chemical stratification in the Orion Bar
in greater detail.

\section{Observations}
\label{sec:observations}

This work is based on data obtained as part of the Spectral Legacy Survey
\citep[SLS,][]{plume2007}, being conducted at the James Clerk Maxwell
Telescope\footnote{The James Clerk Maxwell Telescope is operated by The Joint
  Astronomy Centre on behalf of the Science and Technology Facilities Council of
  the United Kingdom, the Netherlands Organisation for Scientific Research, and
  the National Research Council of Canada.} on Mauna Kea, Hawai'i. The SLS is
performing spectral imaging of 2$\arcmin$$\times$2$\arcmin$ fields in the
direction of the Orion Bar and four other targets. Once completed, the spectral
range covered by the SLS will be 330--362 GHz. This Letter presents the first
results from a selection of molecular lines in the already acquired data: two
\cch\ lines and one transition each of SO, \hhco\ and \thco\ between 330 and 350
GHz, and two lines in auxiliary data in the 230 GHz window: \eico\ 2--1 and HCN
3--2 (Table \ref{t:lines}).

The SLS uses the 16-pixel HARP receiver (325--375\,GHz) and the ACSIS
correlator. The angular resolution of the JCMT at the observed frequencies is
14--15$\arcsec$ ($\approx$30\,mpc at the distance of the Orion Bar); the
\texttt{harp4\_mc} jiggle position switch mode produces 2$\arcmin$$\times$2$\arcmin$ maps
sampled every 7\farcs5. The spectra are calibrated by observations at an off-position 10$\arcmin$ to the southeast. At an exposure time of 4 minutes per map pixel, the rms
noise level in the current SLS data is 0.07--0.08 K in velocity bins of 0.83--0.89\,\kms. Single sideband system temperatures for these observations are $\sim$700\,K.

The HCN 3--2 and \eico\ 2--1 lines were observed in 2005 with the A-band
(215--275\,GHz) receiver and the DAS spectrometer. The same
2$\arcmin$$\times$2$\arcmin$ field as with HARP was mapped at half-beamwidth
spacing, using the raster mapping mode. The JCMT beam size at 220--266 GHz is 19--23$\arcsec$ ($\approx$40\,mpc). The off-position for the A-band observations was at 20$\arcmin$ southeast of the map center. The HCN and \eico\ observations reach a noise level of
0.5--0.6~K at a spectral resolution of 0.18 and 0.21\,\kms, respectively. System temperature
values are $\sim$350\,K (DSB); the integration time is 19 seconds per map pixel. The pointing accuracy was checked every hour and was generally better than 2\arcsec\ during all observing runs.

The data were manually reduced using standard \texttt{Starlink} procedures. 
The SLS observations were handled in blocks of 1.6 GHz effective bandwidth and the
other data as one set per emission line.
The time series were converted to data cubes and first-order polynomial
baselines were subtracted. Line intensities quoted here are
given in terms of the main-beam temperature $T_\mathrm{mb}$, i.e., they have
been corrected for the JCMT main-beam efficiency of 63\% in the 345 GHz window
and 69\% in the 230 GHz window. Single-channel signal-to-noise ratios at the peak in the northeast slice are $\sim$16 for \eico\ and SO, $\sim$40 for HCN, \hhco\ and \cch, and 900 for \thco.

\section{Results}
\label{sec:results}

\begin{figure}
  \resizebox{\hsize}{!}{\includegraphics[angle=270]{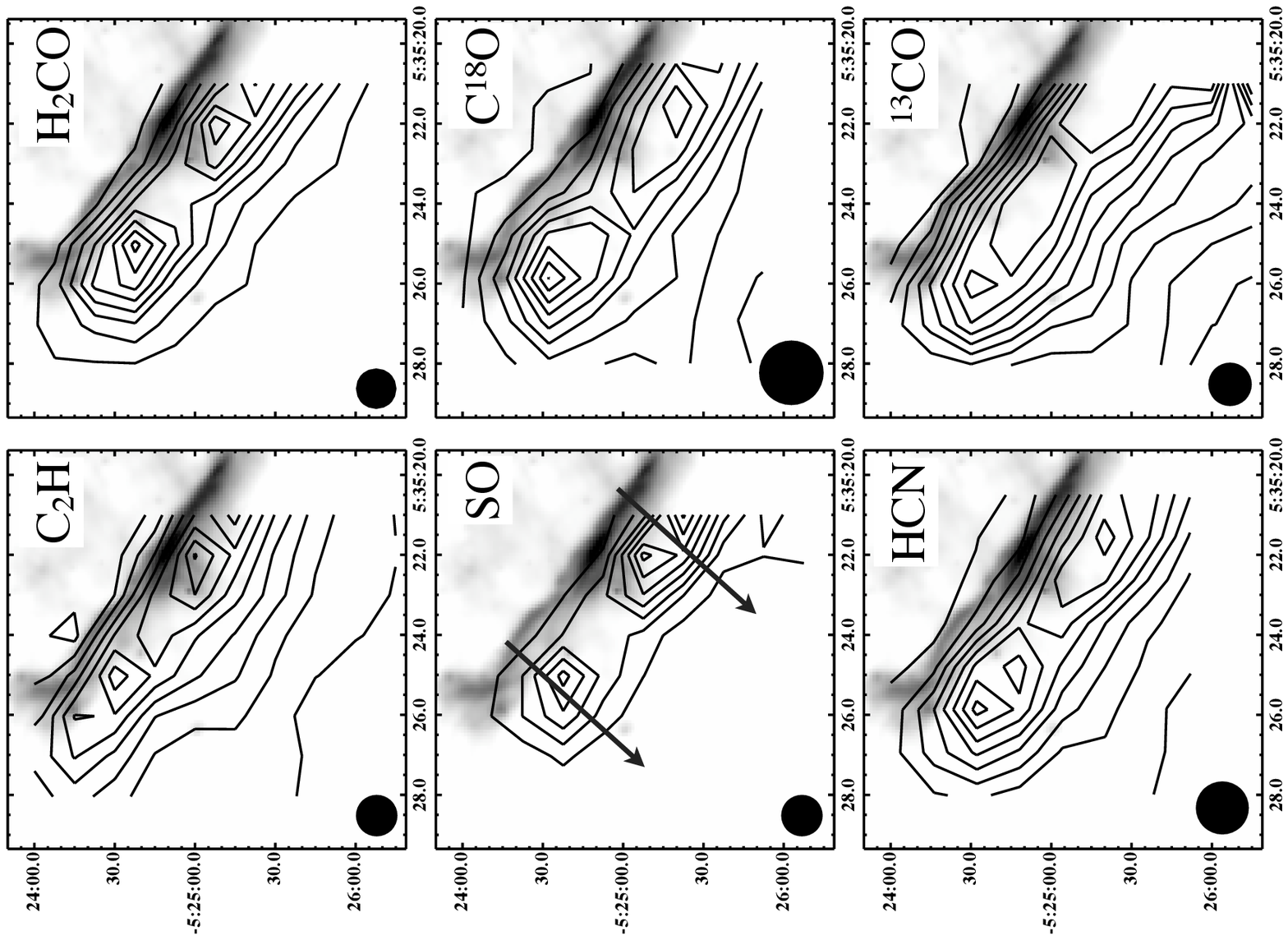}}
  \caption{Images of molecular emission, observed with the JCMT toward the Orion
    Bar of the species and transitions listed in Table~\ref{t:lines}. Line
    intensities are integrated over 5--8\,\kms, depending on the line width. The
    \cch\ transition shown here is the 4$_{9/2}$--3$_{7/2}$ transition; the
    other transition shows a similar spatial distribution. Contour levels are
    drawn at 10, 20, \ldots, 90\% of the maximum integrated intensity for every
    map: 13.3\,\kkms\ for \cch, 8.7 for \hhco, 7.7 for SO, 19.9 for \eico,
    70.5 for HCN, and 276.8 for \thco. Note that the measured maximum $\int
    T_\mathrm{mb} \mathrm{d}v$ values across the northeast slices (Table~\ref{t:lines})
    are not the same as the highest value across the entire map. The grayscale
    background image shows \textit{Spitzer Space Telescope} 8\,$\mu$m continuum
    emission. The filled circle in each frame indicates the beam size at the relevant frequency. 
    Axes are annotated with right ascension and declination (J2000).}
  \label{fig:images}
\end{figure}

Figure~\ref{fig:images} presents images of the observed molecular emission,
overlaid on an 8\,$\mu$m image of the Orion Bar taken with the \textit{Spitzer
  Space Telescope} (Megeath et al., in prep.). The mid-infrared emission traces warm dust at the surface layer between the \HII region and the
molecular cloud. The molecular emission is seen to peak to the southeast of this
layer, and the JCMT data reveal a clear stratification. The \cch\ emission peaks
close to the surface layer, followed by SO and \hhco, whereas CO and HCN peak
deep into the molecular cloud.

\begin{figure}
  \resizebox{\hsize}{!}{\includegraphics{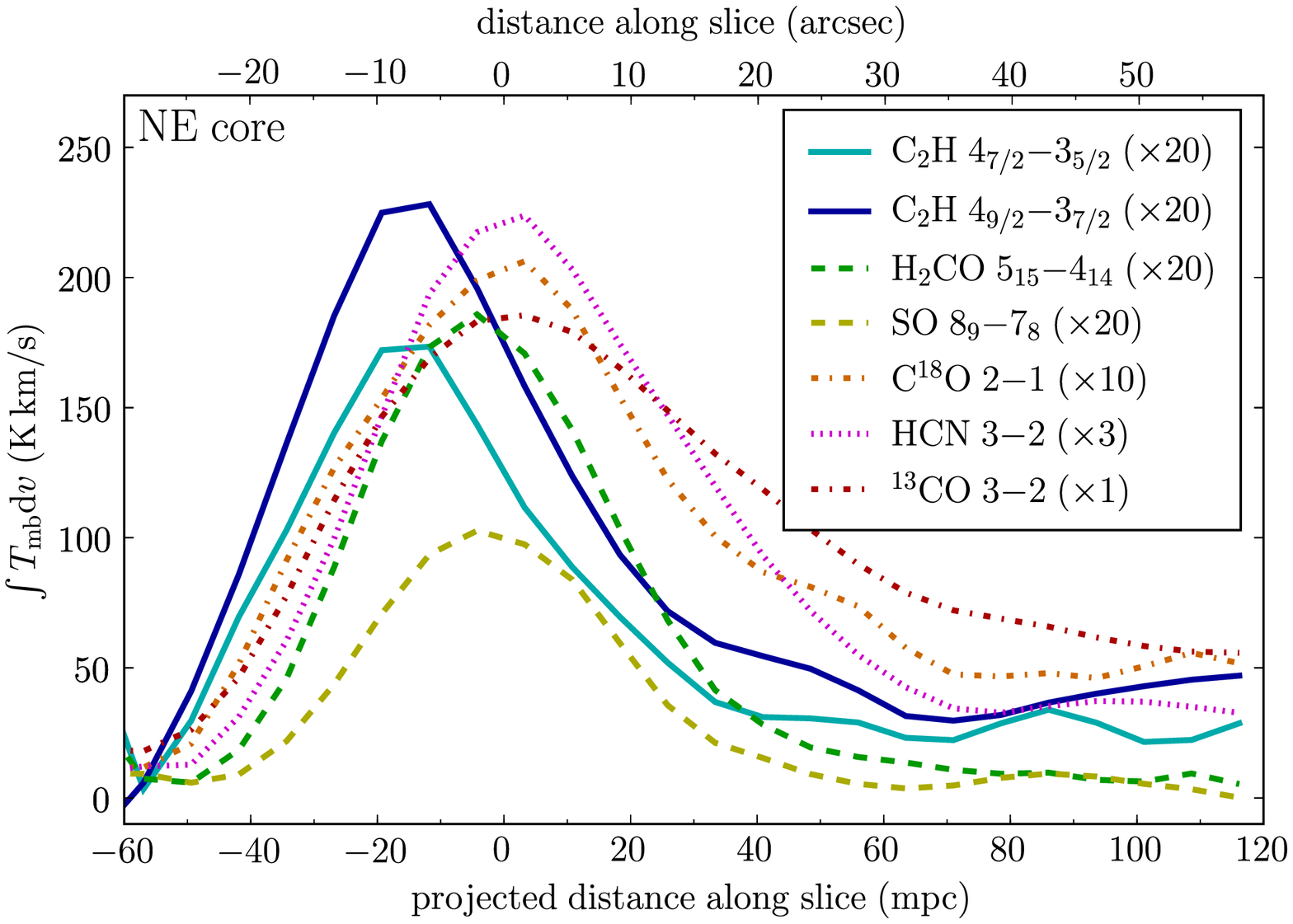}}
  \resizebox{\hsize}{!}{\includegraphics{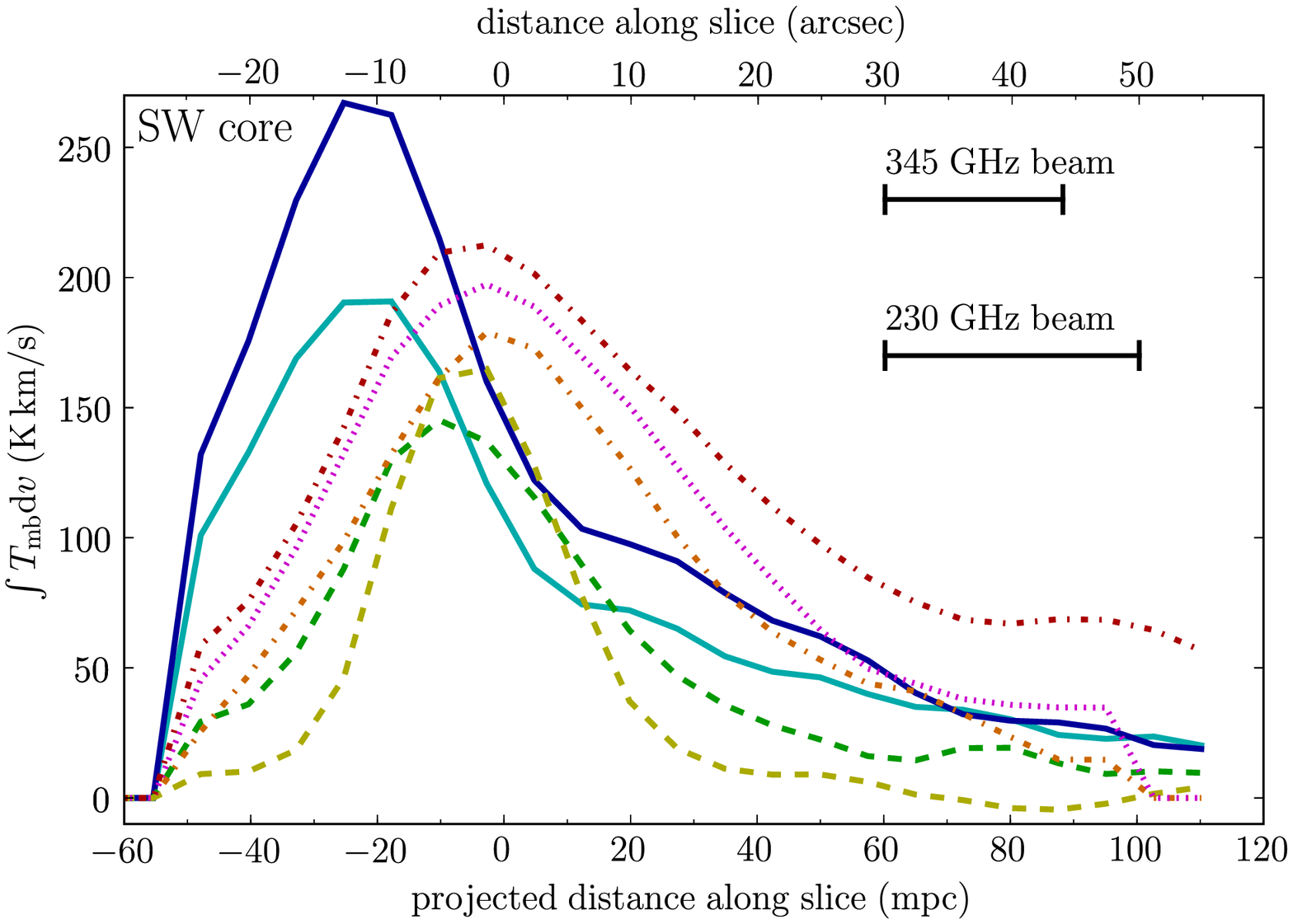}}
  \caption{Profiles of molecular emission along slices perpendicular to the
    ionization front at position angle 225\degr (see arrows in Fig.~\ref{fig:images}): (top panel) through
    the northeast core; (bottom panel) through the southwest core. The zero
    point of the distance scale is placed at the \eico\ peak position: (RA$=$$5^\mathrm{h}35^\mathrm{m}25\fs4$, $\delta$=$-5\degr24\arcmin37\arcsec$) for the northeast slice and (RA$=$$5^\mathrm{h}35^\mathrm{m}21\fs6$, $\delta$=$-5\degr25\arcmin19\arcsec$) for the southwest slice.}
  \label{fig:slices}
\end{figure}

The stratification is more easily visible in Fig.~\ref{fig:slices} which shows
slices through the images in Fig.~\ref{fig:images}, taken along the arrows in
Fig.~\ref{fig:images}, perpendicular to the ionization front. To produce these intensity profiles, all maps are regridded to 3\farcs75 pixels, roughly half the pixel size of the original maps. Fitted peak positions and peak fluxes for the seven profiles are listed in Table~\ref{t:lines}, along with the spectroscopic parameters of the observed lines; the position uncertainties include Gaussian fitting errors ($< 1\arcsec \approx 2$\,mpc).

In addition to the stratification, the molecular emission shows a two-peak
structure. The northeast peak is brighter in HCN and \hhco, 
while the southwest peak is brighter in SO and the peak brightness is roughly equal for CO
and \cch. These variations are probably caused by variations in the filling factor of dense gas, as seen, e.g., in the high-resolution data by \citet{lis2003}. The measured line widths of \mbox{5--8\,\kms}\ will be useful
for future detailed modeling, however, this paper focuses on integrated intensities.

We use the observed line intensities and the radiative
transfer program \texttt{RADEX} \citep{vandertak2007} to estimate beam-averaged molecular
column densities. Although temperatures are
known to range from 50\,K in dense clumps up to 150\,K in the interclump medium
\citep[][]{batrla2003,lis2003}, we assume a kinetic temperature of 85\,K and a
volume density of 10$^5$\,\ccm, as applicable to the extended molecular gas
\citep{hogerheijde1995}. Since some of the line emission may originate from
dense clumps, we also list column densities estimated in the high-density (LTE)
limit (Table~\ref{t:lines}). Both for the $10^5$\,\ccm\ case as well as the LTE limit, varying the temperature by $\pm$30\,K changes the
column densities by $\lesssim$30\% (where a higher temperature gives a lower
column density), so the molecular column densities are reasonably well
constrained.

Our values for the column densities agree with previously published values to a factor of $\sim$2: \citet{hogerheijde1995} for \hhco, \eico\ and HCN, \citet{jansen1995} for \cch, and \citet{leurini2006} for SO. 
The published column densities for HCN and SO are closest to our high-density (LTE) estimates
(Table~\ref{t:lines}); this is consistent with these two species being confined
to the high-density gas while the \hhco\ emission has contributions from both
density phases.
The spatial distribution of the emission in this work is similar to that of previously observed single-dish emission from molecules such as $^{12}$CO, \thco, HCO$^+$, H$_2$, CN and CS \citep[e.g.,][]{tauber1994,vanderwerf1996,simon1997}.

To convert the observed \eico\ line intensity to a total \hh\ column density, we
assume an isotopic ratio $^{16}$O/$^{18}$O=500 \citep{wilson1994} and a CO
abundance of \pow{1.1}{-4} as applicable for the Orion Bar PDR \citep{johnstone2003b}. The result is $N$(\hh) = \pow{1.0}{23}\,\scm, which implies the following molecular
abundances at the peak locations: $x$(\cch) = \pow{2}{-9}; $x$(\hhco) = \pow{4}{-10}; $x$(SO) = \pow{7}{-10}; 
$x$(HCN) $>$ \pow{5}{-10}. Here we have assumed that the emission from all
molecules except CO originates in the high-density gas. 
Based on \citet{jansen1995}, who propose the clumps to encompass 10\% of the material along every line of sight, we estimate the clumps to provide for roughly half of the total column density for those molecules that exist in both phases. Hence, the resulting uncertainty in the derived abundances would not exceed a factor 2.
Since \cch\ and CO are hardly coexistent, the \cch\ abundance is probably underestimated by a factor of 3--10.

\begin{table*}
\begin{minipage}[t]{\textwidth}
\caption{Observed molecular lines.} 
\label{t:lines}
\begin{tabular}{ll@{\ }r@{\ }l@{\ }ll c r@{\ }l r@{\ }l c r@{\ }l cc}
\noalign{\smallskip}
\hline
\hline
\noalign{\smallskip}
Molecule &\multicolumn{4}{c}{Transition} & Frequency &$E_{\rm up}$& \multicolumn{5}{c}{Peak position (mpc)$^\mathrm{(d)}$}  &  \multicolumn{2}{c}{$\int T_{\rm mb}{\rm d}v$$^\mathrm{(e)}$} & \multicolumn{2}{c}{\nmb\ (\scm)$^\mathrm{(g)}$} \\
       &   &     & &       &  (MHz)    &  (K)  &  \multicolumn{2}{c}{NE slice}      & \multicolumn{2}{c}{SW slice}  & model & \multicolumn{2}{c}{(\kkms)}         & $n$=10$^5$\,\ccm & LTE \\
\noalign{\smallskip}
\hline
\noalign{\smallskip}
\cch    &$N_J$        &=& $4_{7/2}$&$\to$ $3_{5/2}$ &349400.5$^\mathrm{(a)}$ &42  & -14 & $\pm$ 1.8 & -23 & $\pm$ 1.0 & -16 & 8.6 & $\pm$ 1.6 & \pow{4.4}{14} & \pow{2.3}{14} \\
\cch    &$N_J$        &=& $4_{9/2}$&$\to$ $3_{7/2}$ &349337.8$^\mathrm{(b)}$ &42  & -14 & $\pm$ 1.5 & -23 & $\pm$ 1.6 & -16 & 11.3 & $\pm$ 2.0 & \pow{4.9}{14} & \pow{2.4}{14} \\
o-\hhco    &$J_{KpKo}$    &=& $5_{15}$ &$\to$ $4_{14}$ &351768.6     &62  & -3 & $\pm$ 1.3 & -7 & $\pm$ 1.9 & -12 & 9.3 & $\pm$ 1.6 & \pow{7.7}{14} & \pow{3.8}{13} \\
SO    &$N_J$        &=&  $8_9$   &$\to$ $7_8$    &346528.5     &79  & -2 & $\pm$ 1.4 & -5 & $\pm$ 1.0 & 8 & 5.2 & $\pm$ 0.9 & \pow{1.4}{15} & \pow{7.2}{13} \\
\eico    &$J$          &=&  2     &$\to$ 1       &219560.4     &16  &  0 & $\pm$ 1.2 & 0 & $\pm$ 1.4  & 0 & 20.3 &$\pm$ 3.4 & \pow{2.3}{16} & \pow{2.9}{16} \\
HCN    &$J$          &=&  3     &$\to$ 2       &265886.2$^\mathrm{(c)}$ &26  & 3 & $\pm$ 2.5 & -1 & $\pm$ 1.8  & 8 & 74.0 &$\pm$ 14 & $>$\pow{6.5}{12}$^\mathrm{(f)}$ & $>$\pow{4.9}{13}$^\mathrm{(f)}$ \\
\thco    &$J$          &=&   3     &$\to$ 2       &330588.0     &32  & 4 & $\pm$ 1.9 & -1 & $\pm$ 3.0  &  2 & 186 &$\pm$ 32 & $>$\pow{2.1}{16}$^\mathrm{(f)}$ & $>$\pow{2.5}{16}$^\mathrm{(f)}$ \\
\noalign{\smallskip}
\hline
\end{tabular}\\
(a): Blend of $F$=4-3 and $F$=3-2 hyperfine components (separation 1.4\,MHz). 
(b): Blend of $F$=5-4 and $F$=4-3 hyperfine components (separation 1.3\,MHz). 
(c): Blend of $F$=3-3, $F$=3-2 and $F$=2-1 hyperfine components (total separation 1.5\,MHz). 
(d): Fitted position of the emission peak along the slice through the northeast and southwest cores (see Fig.~\ref{fig:images}), and the modeled slice (Fig.~\ref{fig:model}, bottom panel), all relative to the \eico\ peak.  
(e): Fitted value of integrated intensity at peak position along the northeast
slice (see Fig.~\ref{fig:images}); uncertainties include fitting uncertainty and
15\% absolute calibration uncertainty. 
(f): Column density uncertain due to high optical depth; values listed here are lower limits.
(g): Beam-averaged column densities for $n$=10$^5$\,\ccm\ and in LTE, both at $T$=85\,K. 
\end{minipage}
\end{table*}

\section{Discussion}
\label{sec:disc}

\begin{figure}
  \resizebox{\hsize}{!}{\includegraphics{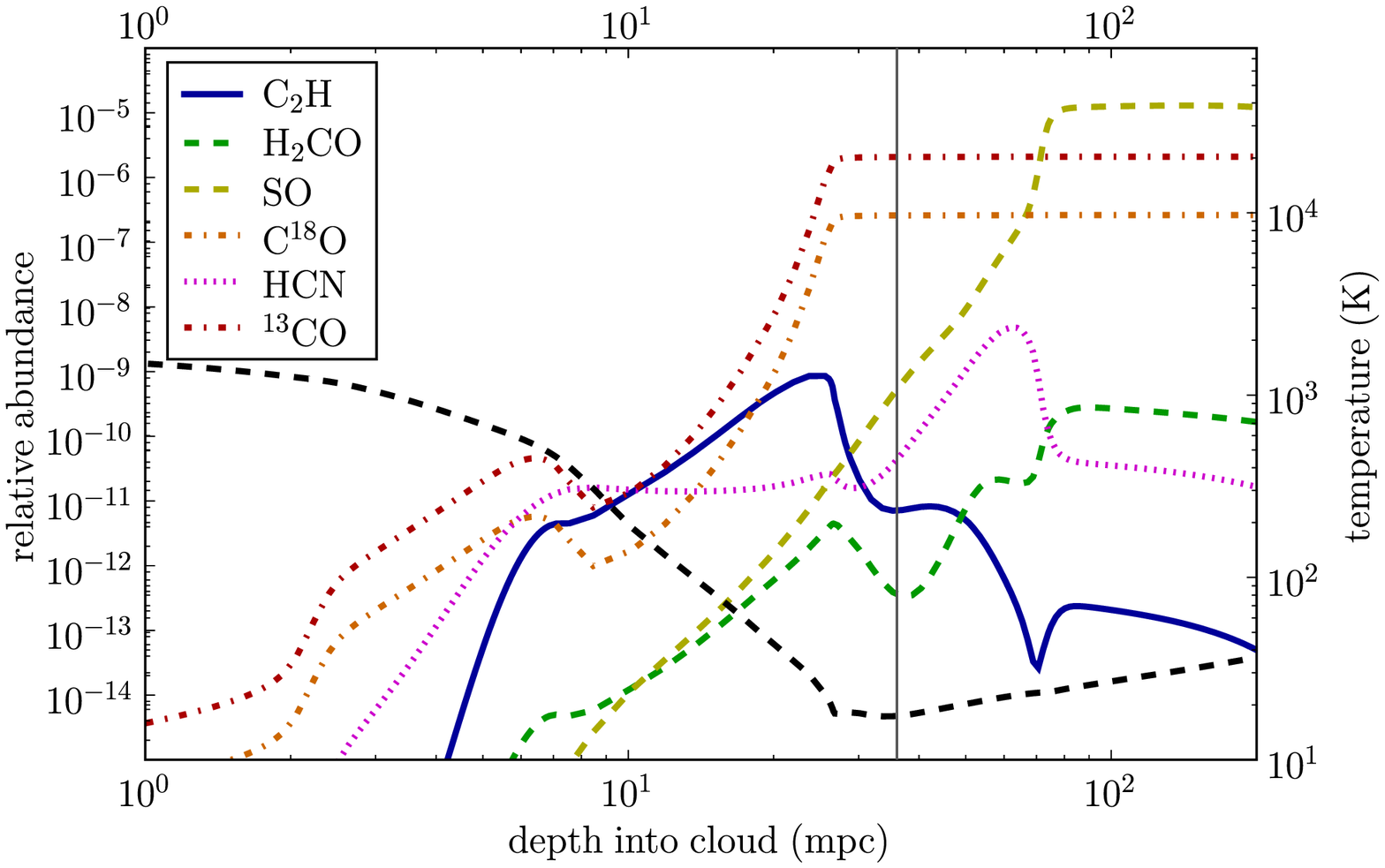}}
  \resizebox{\hsize}{!}{\includegraphics{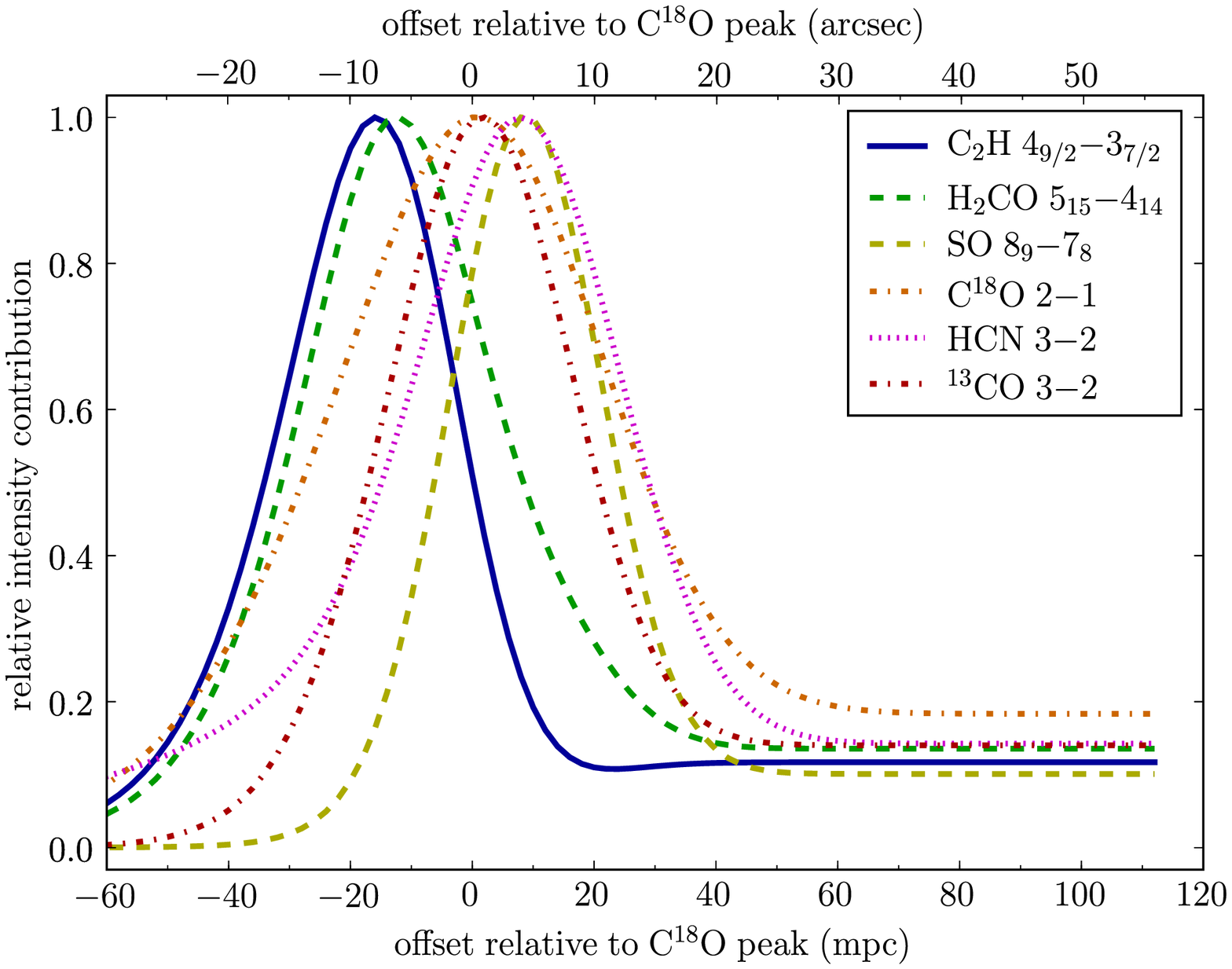}}

  \caption{Top panel: model calculation of relative abundances as a function of
    depth into the cloud. The dashed black curve shows the temperature profile. 
    The thin vertical line indicates the zeropoint of the horizontal scale in the bottom panel.
    Bottom panel: modeled relative intensity profiles in the direction perpendicular to the line of sight,
    convolved by the JCMT beam. In this panel, the zeropoint of the depth
    scale is set to the modeled \eico\ peak, at 36\,mpc from the edge of the molecular cloud.}
  \label{fig:model}
\end{figure}

To interpret our observations we compare them to the predictions from a PDR
model. Previous models of the Orion Bar do not present abundances for all six species in our data, so we use
the global conditions of the Bar ($\chi$=\pow{3}{4}\,$\chi_0$, $n_{\rm H}$=$10^5$
cm$^{-3}$, see Sect.~\ref{sec:intro}) to construct a simple model using the KOSMA-$\tau$
PDR code \citep{rollig2006}. We refer to \citet{rollig2007} for a comparative study of other PDR codes. To simulate the plane-parallel geometry of the Orion Bar we represent it by one side of a very large sphere. Starting from atomic abundances of $x$(O)$=$\pow{3.0}{-4}, $x$(C)$=$\pow{1.4}{-4}, $x$(S)$=$\pow{2.8}{-5} and $x$(N)$=$\pow{1.0}{-4} \citep[see also][Table~3]{cubick2008}, the model
computes the equilibrium chemistry and temperature of the cloud using reaction rates from the UMIST database \citep{woodall2007}. Based on the geometry in Fig.~1 of \citet{jansen1995} we simulate the PDR as a layer of thickness 49\,mpc where the Bar represents a part of the layer that is
parallel to the line of sight, so that its apparent depth is increased by a
factor 10 compared to its thickness. 
We note that the critical densities are $\sim$$10^7$\,cm$^{-3}$ for all transitions in Table~\ref{t:lines} except those of the CO isotopes, while the highest \hh\ density reported by \citet{lis2003} is also $10^7$\,cm$^{-3}$. Hence, to approximate the spatial distribution of the line emission,  we assume optically thin LTE and weight the abundance profiles with a Boltzmann cutoff: $I_{\rm rel} = (x/x_0)\,{\rm exp}{ (1-T_{\rm gas} / E_{\rm up} ) }$ if $T_{\rm gas}< E_{\rm up}$ and $I_{\rm rel} = x/x_0$ otherwise. Here $I_{\rm rel}$ is a normalized intensity, $x$ is abundance, $T_{\rm gas}$ is the modeled local gas temperature (see Fig.~\ref{fig:model}, top panel), $E_{\rm up}$ is given in Table~\ref{t:lines} and the normalization factor $x_0$ is chosen from line to line.
The distribution is convolved with the JCMT beam at the appropriate frequencies to obtain the predicted relative intensity profiles (Fig.~\ref{fig:model}, bottom panel).

Table~\ref{t:lines} compares modeled positions of the peak emission to fitted
peak positions in the observed profiles for each species. Overall
our simple PDR model matches the observed layering of \cch-\hhco-CO.
For \hhco\ the highest abundance occurs in the cold shielded material, but the warm outer layers dominate the emission due to the higher $E_\mathrm{up}$ for the observed transition.

There are also significant discrepancies between the model and the observations. (i) The model predicts that SO peaks deep in the cold material (at 8\,mpc), following the distribution of HCN, not at the location where \hhco\ peaks (in front of CO) as is observed. 
(ii) Conversely, the \hhco\ peak in both slices is observed closer to the peaks of the CO isotopes than to the \cch\ peak, while the model predicts \hhco\ to peak significantly closer to \cch.
(iii) For HCN the model peaks 5--9\,mpc deeper than is observed and
(iv) the observations show an HCN emission wing toward the PDR surface which is weaker in the model. 
(v) We find that the absolute abundance of \hhco\ and the lower limit to the absolute abundance of HCN at the model peak are consistent with the observations, while the model abundance of \cch\ is $\sim$2 times smaller than observed. 
(vi) In addition, our modeled SO abundance is orders of magnitude larger than the observed value.

The underestimate of the \cch\ abundance predicted by our pure gas-phase model might be explained by \cch\ being formed via photo-destruction of PAH molecules, as proposed by \citet{pety2005}. This explains a detectable \cch\ abundance in the outer layers where the PAH emission peaks \citep{vanderwerf1996}.

Our modeled SO abundance is a factor \pow{1.4}{4} larger than observed. This is explained by the too high gas phase sulfur abundance of \pow{2.8}{-5} used in the KOSMA-$\tau$ model, which represents diffuse clouds and no depletion \citep{federman1993}.
Recent estimates of cosmic sulfur abundance provide values as low as \pow{1.4}{-5} \citep{asplund2005} and sulfur depletion factors in dense clouds may range
from 4 \citep{goicoechea2006} to 1000 \citep{tieftrunk1994}. Depletion factors of several $10^3$ would be required to match the Orion Bar observations with the current model. The low depletion seems not to apply to the entire Orion region, but to be specific for the Horsehead, which has a particularly low density.

The abundance of \hhco\ derived in the LTE limit is slightly higher than the modeled abundance; the discrepancy increases sharply as soon as the \hhco\ gas is not completely in LTE.  
This abundance discrepancy might be explained by the evaporation of
molecules from warm ($\sim$30\,K) dust grains. Based on reaction rates from
\citet{woodall2007}, the photo-dissociation lifetime for \hhco\ at
$A_\mathrm{V} = 5$\,mag is only $\sim$10\,yr, which introduces the need for a
continuous supply of \hhco, e.g., from grain surfaces \citep{herbst2000}. This is consistent with the suggestion by  \citet{leurini2006} that \hhco\ may originate primarily from the warm interclump medium.

Finally, the observed broad spatial profiles for HCN and \thco\ (Fig.~\ref{fig:slices}) may be explained by
optical depth effects, since the optical depth inferred by \texttt{RADEX} is 5--10
for these molecules and $<$1 for the others. The different spatial distributions
of the two CO isotopologs result mainly from the different beam sizes.

Future observations will test our hypothesis that grain surface chemistry
plays a key role in PDRs. One such test will be JCMT-SLS observations of \meth\
in the Orion Bar: while \citet{leurini2006} suggest that \meth\ traces the
clumps and \hhco\ traces the interclump medium, both molecules are thought to be
the products of CO hydrogenation on ice \citep{watanabe2004}, so their spatial
coincidence would mark a critical validation step. 
Other analysis will explore sulfur chemistry using sulfur-bearing species covered by the SLS, such as CS and SO$_2$.
Once completed, the SLS data set
will enable studies of multiple transitions of various molecules, so that the
excitation of molecules (as well as their velocity structure) can also be
compared to models. In the longer term, (sub)millimeter-wave imaging
spectroscopy of other Galactic PDRs is needed to establish how common grain
chemistry is and how it depends on environment. Important tests will be spectral
surveys with the HIFI instrument on board ESA's \textit{Herschel} Space
Observatory, which will target hundreds of molecular lines in many Galactic PDRs
and allow a detailed picture of their chemistry to emerge.

\begin{acknowledgements}
  The authors acknowledge the JCMT staff for their support, Markus R\"ollig for help with the PDR model, and Peter Schilke, Maryvonne Gerin, and an anonymous referee for their careful reading of the manuscript.
\end{acknowledgements}

\bibliographystyle{aa}  
\bibliography{../../literature/allreferences}

\end{document}